# Polymer Nanofibers with Outstanding Thermal Conductivity and Thermal Stability: *Fundamental Linkage between Molecular Characteristics and Macroscopic Thermal Properties*

Teng Zhang,[†] Xufei Wu,[†] Tengfei Luo[†,‡,*]

[†]Aerospace and Mechanical Engineering and [‡]Center for Sustainable Energy at Notre Dame, University of Notre Dame, Notre Dame, Indiana 46556, United States

## Abstract

Polymer nanofibers with high thermal conductivities and outstanding thermal stabilities are highly desirable in heat transfer-critical applications such as thermal management, heat exchangers and energy storage. In this work, we unlock the fundamental relations between the thermal conductivity and thermal stability of polymer nanofibers and their molecular characteristics by studying the temperature-induced phase transitions and thermal transport of a series of polymer nanofibers. Ten different polymer nanofibers with systematically chosen molecular structures are studied using large scale molecular dynamics simulations. We found that high thermal conductivity and good thermal stability can be achieved in polymers with rigid backbones, exemplified by **π**-conjugated polymers, due to suppressed segmental rotations and large phonon group velocities. The low probability of segmental rotation does not only prevent temperature-induced phase transition but also enables long phonon mean free paths due to reduced disorder scattering. Although stronger inter-chain interactions can also improve the thermal stability, polymers with such a feature usually have heavier atoms, weaker backbone bonds, and segments vulnerable to random rotations, which lead to low thermal conductivities. This work elucidates the underlying linkage between the molecular nature and macroscopic thermal properties of polymer nanofibers, which is instrumental to the design of thermally conductive polymer nanofibers with high temperature stabilities.

* Tel.:+1 574 631 9683; E-mail: tluo@nd.edu. (Corresponding Author)

# 1. Introduction

Amorphous polymers, such as polymer foam and composite polymer, are usually used as thermal insulator due to their low thermal conductivities (~O(0.1) W/mK).[1] However, contrary to common wisdom, recent studies have shown that polymers such as polyethylene (PE) chains are not intrinsically of low thermal conductivities.[2, 3] Using molecular dynamics (MD) simulations, PE single chains were predicted to have very high conductivity (~O(100) W/mK).[2] While a single molecular chain in a strictly straight geometry simulated in Ref. [[2]] is impractical in nature, their work shed the light on possible large thermal conductivity in highly aligned polymer fibers, in which the molecular chains are very straight. From single chain to PE fibers, the impact of van der Waals (vdW) forces between polymer chains were also studied by MD simulations.[3] Although the thermal conductivity of PE bulk crystal was found to be reduced to ~45 W/mK due to the vdW-force-induced anharmonic phonon scattering,[3] this value is still larger than many metals.

PE fibers with high thermal conductivity were also reported in experimental studies.[4-8] Oriented PE fibers were fabricated by mechanical drawing process, and the thermal conductivity was improved from ~0.55 W/mK to 18.8 W/mK.[4] PE nanowire arrays of high thermal conductivity (up to ~20 W/mK) were also fabricated by nanoporous template wetting technique,[5] and the ~20 W/mK value for PE nanofiber array was also confirmed by recent time-domain thermoreflectance (TDTR) measurements.[6] High thermal conductivity up to 42 W/mK was also reported in high modulus gel-spun PE nanofibers,[7] but this result may suffer from errors due to thermal radiation near room temperature. A very high thermal conductivity of 104 W/mK measured using cantilever method was reported for ultra-drawn PE fibers recently,[8] leading to the discussion on the upper-limit of polymer thermal conductivity. The availability of thermally conductive polymers can expand the plastic industry, replacing metals and ceramics in heat transfer devices and equipment, leading to energy and cost savings.

The thermal conductivity of polymers is also found to be significantly morphology-dependent.[9-14] From amorphous phase to crystalline structures, the thermal conductivity of PE was predicted to increase from ~0.3 W/mK[15] to ~50 W/mK by MD simulations.[9] Within the crystalline phase, a phase transition of PE at ~400 K can destroy the along-chain segmental order, leading to an abrupt decrease in the thermal conductivity by almost one order of magnitude (MD results).[9, 10] Complex phase transition phenomena in PE nanofiber were observed experimentally,[16] and the corresponding thermal conductivity reduction was also reported by TDTR measurements.[6] Such morphology changes do not necessarily destroy the entity of the polymer nanofibers but significantly impair the thermal conductivity, presenting thermal stability issues in thermal transport. This limits the application of thermally conductive PE nanofibers at high temperatures. It is thus imperative to take the thermal stability into consideration when choosing or designing polymer nanofibers for heat transfer applications.

Experimental efforts in improving thermal conductivity of polymers have been focused almost exclusively on compositing.[17-27] Changing the morphology to enhance thermal conductivity has been studied for a few different polymers by drawing them into nanofibers.[6, 8, 28, 29] However, the selection of thermally conductive nanofibers has not been done under theoretical guidance. Wang *et al.* made a connection between the modulus of polymer nanofibers and their thermal conductivities, but the relation is not monotonic.[6] On the other hand, thermal stabilities of the polymer nanofibers are largely ignored when exploring high thermal conductivities. It is thus highly desirable that we can tell whether a polymer has the potential to become thermally conductive with high temperature stability when formed into nanofibers by simply looking at the chemical composition of the molecules. This ability would be instrumental to the design of thermally conductive polymer nanofibers for applications at different temperature ranges. This calls for the elucidation of the fundamental linkage between the thermal properties and the molecular characteristics of polymers.

In this work, we use large scale molecular dynamics (MD) simulations to study the thermal transport in a variety of polymer nanofibers as a function of temperature which influences their morphologies. It is worth noting that classical MD simulations do not include any quantum effect, and all vibrational modes are excited regardless of the temperature in the simulations. In reality, some high frequencies modes are actually not excited when the temperature is lower than the Debye temperature. However, all the simulations presented in this work are at 300K or higher, and the low frequency acoustic modes, which are believed to be the most important contributors to thermal transport, are excited in this temperature range in reality. High frequency modes, especially those related to the light hydrogen atoms, are not excited at room temperatures. However, these modes are usually highly localized and do not contribute much to the thermal transport.[30-33] The simulation details are discussed in the simulation section. We first relate the morphology-influenced thermal conductivity to the inter-chain and intra-chain interatomic interactions through parametric studies of a model PE system. The obtained relations are then generalized by analyzing the temperature-dependent thermal conductivities of ten different polymers nanofibers featuring weak and strong inter-chain and intra-chain interactions. Finally, we analyze the fundamental relation between the molecular characteristic and phonon properties in these polymer nanofibers. Based on these analyses, we are able to provide a theoretical connection between the thermal properties of polymer nanofibers and their molecular features.

## 2. Results and Discussions

### 2.1. Parametric Study of Thermal Transport in PE

Although PE has high thermal conductivity (~50 W/mK) at 300K, a sharp drop to ~5 W/mK at ~400K due to a phase transition which leads to random segmental rotations is reported previously (Figure 1).[6, 9, 10, 16] Lindemann's criterion predicts that melting initiates when the amplitude of vibration becomes large enough for adjacent atoms to occupy the same space.[34-36] This means that if the atomic movements can be confined, the phase transition can be suppressed. Generally, the two dominant factors that influence the large motions of atoms in a polymer chain are: dihedral angle strength and inter-chain interaction, because they are usually the weakest interactions among the different interatomic interactions (*e.g.*, bond length and bond angle) in polymer chains. For example, the energy constants of the dihedral angle of PE backbone is 0.2432 Kcal/mole, and the inter-chain energy constant is 0.054 Kcal/mole, but those for bond length and bond angles are 299.67 Kcal/mole and 39.52 Kcal/mole.[37, 38] If the dihedral angle is weak, the segments in the chain can easily rotate and thus lead to disorder along the chain. Such disorder can be suppressed if the inter-chain interaction is strong enough to bundle the chains tightly together and limit the free space for segment rotation.

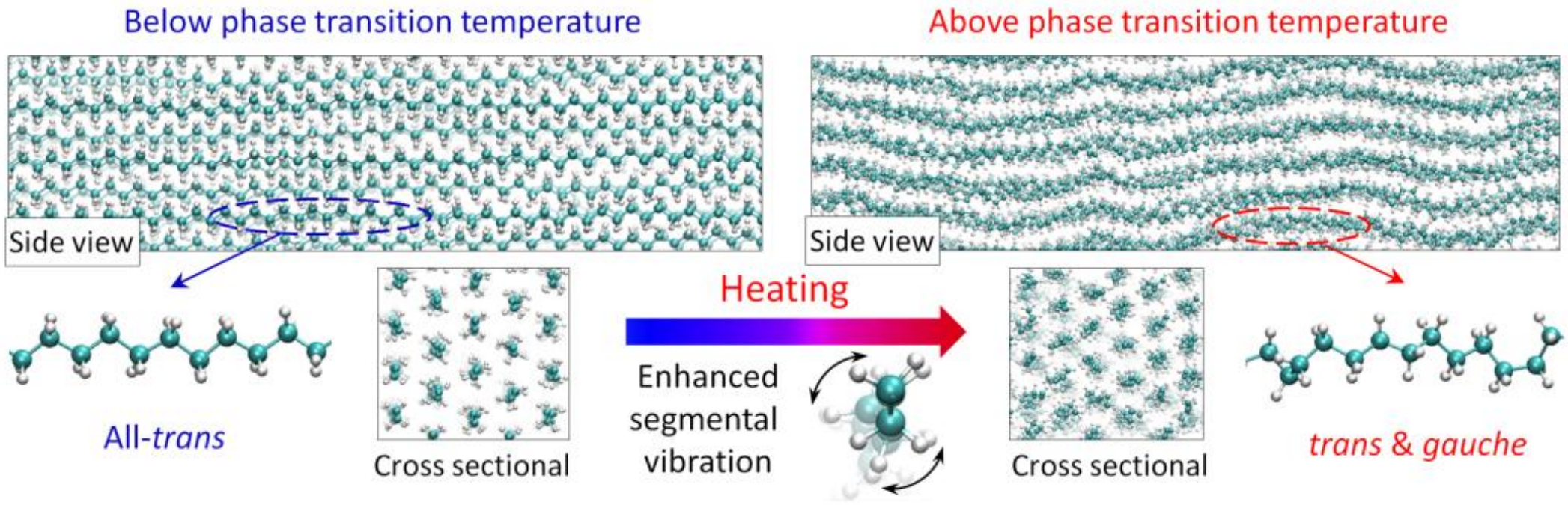

*Figure 1. Phase transition from all-trans conformation crystalline phase to aligned phase with random segmental rotations. When temperature increases above the phase transition temperature, the magnitude of polymer segmental vibration will become large enough to break the inter-chain crystalline lattice, leading to disorders along the polymer chains and dramatically lower thermal conductivity.*

In this part, we perform parametric studies to control the morphology change of a model PE nanofiber. To suppress segmental rotation and confine chain movement, intrinsically stiff chain backbones and stronger inter-chain confinement are artificially realized by increasing the energy constants of the dihedral angle and inter-chain vdW interaction, respectively.[39, 40] In the condensed-phase optimized molecular potentials for atomistic simulation studies (COMPASS) potential,[37, 38] for PE used in this work, the dihedral angle and the 6/9 Lennard-Jones potential terms adopt the following forms:

$$E_{dihedral} = \sum_{n=1}^{3} K_n[1 - cos(n\Phi)] \quad (1)$$

$$E_{Vdw} = \varepsilon \left[2\left(\frac{\sigma}{r}\right)^9 - 3\left(\frac{\sigma}{r}\right)^6\right], r < r_c \quad (2)$$

where $\Phi$ is the dihedral angle value, $K_n$ is the dihedral angle energy constant, $r$ is the distance between two atoms, $\sigma$ is related to atom equilibrium positions, $\varepsilon$ is the energy constant of the vdW interaction, and $r_c$ is the cutoff.

To investigate the relationship between thermal stability and polymer structure, a parametric study on $K_n$ and $\varepsilon$ is performed. With $K_n$ and $\varepsilon$ ranging from 0.8 to 1.2 times the original values, polymer structures from 300K to 500K are characterized,[9, 10, 41] and the corresponding thermal conductivities are calculated (Figure 2). The crystalline PE structures are first prepared by running the simulations in NPT ensembles at 300 K and 1 atm for 1 ns. The structures are then heated up to 500K at a rate of 40 K/ns. Figure 2a shows the simulation volume as a function of temperature for different model systems. We see large jumps in the volumes, which are results from phase transitions. Heating rate is known to influence the phase transition temperatures. In a separate set of simulations, we extract the steady state simulation volumes from equilibrium NPT runs at different temperatures which correspond to infinitely slow heating processes (Figure 2a, black dots). The phase transition temperatures from these two sets of data are within 5 K of each other, showing that the 40 K/ns heating rate does not influence the observed morphology change significantly. Since in this parametric study we are interested in the qualitative relations among interatomic interactions, morphology and thermal conductivity, the morphology characterizations are performed on the structures obtained from the simulations with a 40 K/ns heating rate for convenience.

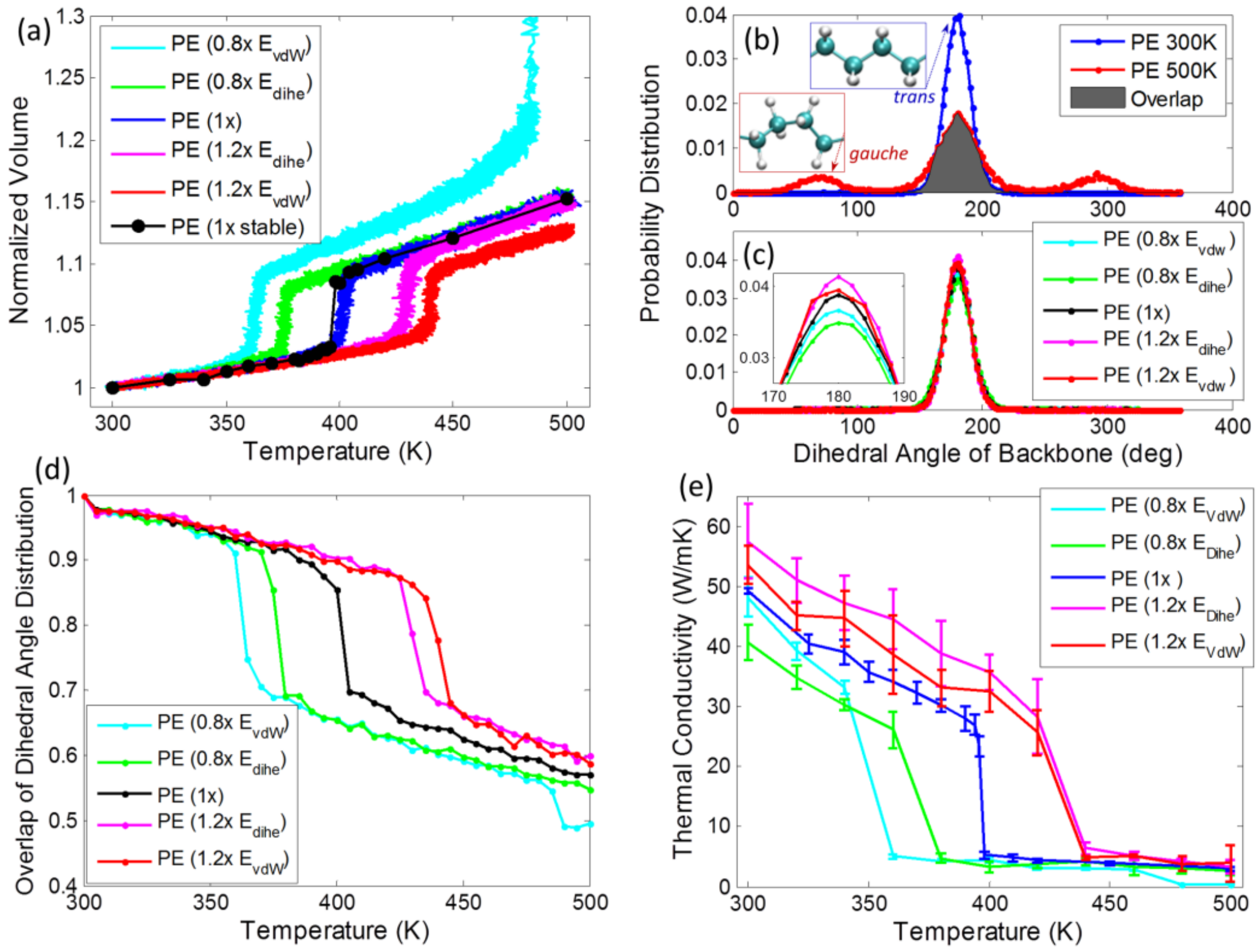

*Figure 2. Morphology-thermal conductivity relation from parametric studies on model PE nanofibers. (a) The volume of the model PE systems with different inter-chain vdW interactions and dihedral energies; (b) normalized dihedral angle distribution of original PE at 300K and 500K, and the overlap of the dihedral angle distributions (grey area); (c) dihedral angle distribution of different model PE nanofibers at 300K; (d) the overlap of dihedral angle distributions from 300K to 500K; and (e) thermal conductivity from 300K to 500K for different model PE nanofibers.*

Figure 2a shows that phase transitions, which are well indicated by the sharp increase in volume, occur at lower temperatures when the dihedral energy or the vdW energy constant is reduced. The phase transition temperatures shift to higher temperatures (up to 440K), when vdW or dihedral energy is made 20% larger than the original values (Figure 2a, pink and red lines). Due to the strong bond and angle interactions in polymers, bond length and angle cannot have large thermal fluctuations.[37, 39] Polymer chain structures in different phases differ mainly due to bond rotation, which can be characterized using dihedral angle distribution.[9-11, 42-44] Below the phase transition temperature, the sharp mono-peak dihedral angle distribution around 180$^o$ indicates all-*trans* conformations of the segments along the chain (Figure 2b, 300K and upper inset). The broadening of the dihedral angle distribution peak around 180$^o$ and the appearance of the *gauche* conformations (dihedral angles of 60$^o$ or 300$^o$) above the phase change temperature indicate coiled chains in which segments rotate more freely (Figure 2b, 500K and lower inset).[9-11]

Larger vdW energy and dihedral energy lead to stronger confinement for chain motion and segment rotation, and thus sharper peaks are observed (Figure 2c, inset). To compare the morphologies and quantify the segmental rotation at different temperatures, we calculate the overlap of the dihedral angle distribution at two difference temperatures. If the structure remains the same, the overlap is 1. When the difference between two structures becomes larger, the overlap of the dihedral angle distribution decreases. Structures at 300 K, which are highly ordered, are set as the reference, and dihedral angle distributions at other temperatures are compared against this reference (*e.g.*, Figure 2b, grey area). In Figure 2d, sharp drops are found in the dihedral angle distribution overlap plot around the phase transition temperatures that are determined from Figure 2a. The much smaller overlaps after the phase transition indicate much more disordered structures, which impair their thermal conductivities due to disorder-phonon scattering.[9-11] Figure 2e shows that the thermal conductivities undergo sharp drops when segmental rotations emerge. The thermal conductivity drops for PE structures with larger vdW and dihedral energies are found to occur at higher temperatures, thus extending the temperature range for high thermal conductivities (Figure 2e). Even at the same temperature, structures with larger vdW and dihedral energies have higher thermal conductivities (Figure 2e), since these structures have less segmental disorders (Figure 2c inset).

From this parametric study, we understand the following principle: stiffer backbones, which are less susceptible to segmental rotations, and stronger inter-chain interactions, which confine the chain movements and thus further suppress segmental rotations, can enable better thermal stability and higher thermal conductivity.

## 2.2. Polymer with Strong Inter-Chain Interactions

To generalize the principle obtained from Section 2.1 to real materials, we first investigate three other polymer nanofibers which have structures similar to PE but with stronger inter-chain interactions: Nylon 6-6, a polyamide class polymer with strong inter-chain hydrogen bonds (H-bond); Teflon, a fluorocarbon solid with large electron dipole on the C-F bond and thus large Columbic inter-chain interactions; and polyketone (PK), which has a strong chain-chain attraction due to the polar ketone groups. The chemical structures of these polymers are shown in Figure 3 together with other polymers studied in this work. To quantify the strength of the inter-chain interactions, the inter-chain energy density is calculated as the total inter-chain pair energy (summation of vdW and Coulombic energies) divided by the total volume. From PE, Nylon, PK to Teflon, a clear increasing trend is found in the inter-chain energy density (Table 1). As the inter-chain energy gets larger, the volume expands much less when heated up from 300K to 500K (Figure 4a).

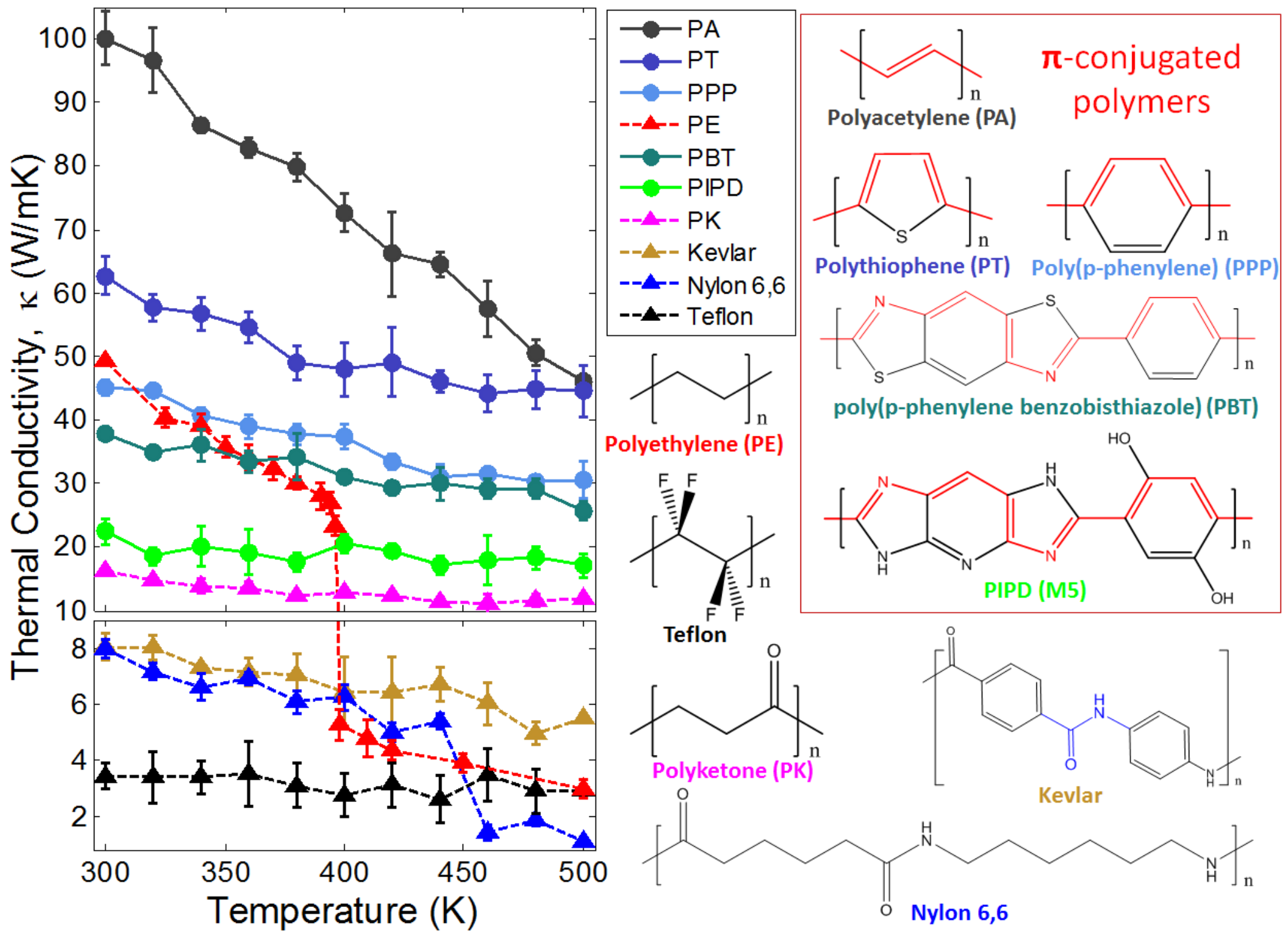

*Figure 3. Thermal conductivity and structure of polymer nanofibers at different temperatures. Π-conjugated polymers are plotted by solid lines with circle markers, while others are plotted by dash lines with rectangular markers.*

Table 1. vdW energy constant ($\varepsilon$) of representative atoms, atomic charge ($q$) and inter-chain energy density of PE, Nylon, PK and Teflon. The atom types are included in the brackets.

| | $\varepsilon$ (Kcal/mol) | $q$ (electron charge) | $q$ (electron charge) | Inter-chain Energy Density (Kcal/Å$^3$/mol) |
|---|---|---|---|---|
| PE | 0.0540 (C) | 0.053 (H) | -0.106 (C) | 0.01 |
| Nylon | 0.2670 (O) | 0.378 (H) | -0.531 (O) | 0.07 |
| PK | 0.2670 (O) | 0.396 (C) | -0.396 (O) | 0.38 |
| Teflon | 0.0598 (F) | 0.500 (C) | -0.250 (F) | 0.74 |

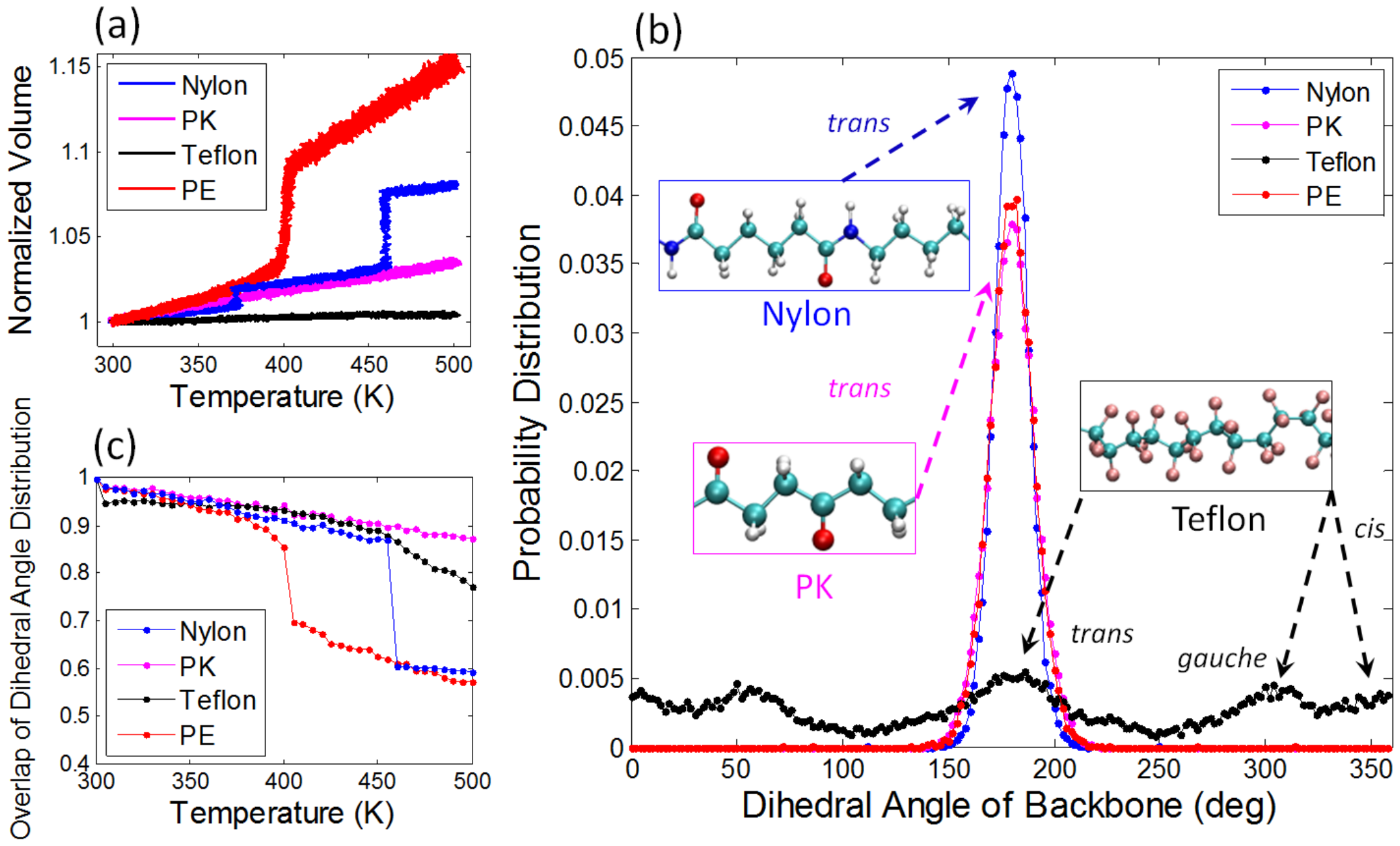

*Figure 4. Structural characteristics of polymer nanofibers with different inter-chain interactions: (a) volume from 300K to 500K, (b) dihedral angle distribution at 300K, and (c) the overlap of dihedral angle distribution from 300K to 500K for PE, Nylon, PK and Teflon.*

To further characterize the polymer structures, dihedral angle distributions at 300K of the four polymer nanofibers are calculated (Figure 4b). Similar to PE, both Nylon and PK have narrow distributions, indicating all-*trans* conformations and straight chains (Figure 4b). For Nylon, segmental rotations are found to emerge above 460K (Figure 4c), and this leads to an abrupt thermal conductivity drop from ~6 W/mK to below 2 W/mK (Figure 3). Experimental data also confirm that phase transition happens around 460K for electrospun Nylon 6-6 nanofibers.[45] It is worth noting that the volume of Nylon shows a discontinuity around 375K (Figure 4a). However, the dihedral angle distribution evolves gradually around this temperature (Figure 4c). It is found that the orientations of some chains in the Nylon nanofiber changed collectively at this temperature, but no random segment rotations along the chains are observed. The thermal conductivity thus shows no significant change around 375K. For PK nanofiber, the good crystalline structure without much segmental rotations is found to be maintained up to 500K (Figure 4a & 4c). As a result, its thermal conductivity only decreases gradually (shown in Figure 3), likely due to anharmonic phonon scattering. A very high melting point of 551K has also been previously reported in experiment for PK.[46] Although Teflon shows little thermal expansion due to the strong inter-chain interactions (Figure 4a & Table 1), the broader and multi-peak dihedral angle distribution even at 300K indicates intrinsic disorder in segmental orientations (Figure 4b). This lack of long range order in Teflon above 300K has been both predicted by simulation and verified by experiments.[47-49] These disorders scatter phonons and they are inherent in Teflon chains even at room temperature. Therefore, Teflon is found to have low thermal conductivity in the entire 300-500K temperature range (Figure 3).

From these four polymer nanofibers, we see that the thermal stability indeed increases as the inter-chain interaction becomes stronger. However, polymers with stronger inter-chain interactions do not necessarily have higher thermal conductivity, which is related to segmental order.

## 2.3. $\pi$-conjugated Polymers with Rigid Backbones

Due to the outstanding electronic-photonic properties and thermochemical stability, $\pi$-conjugated materials are applied in many areas, including chemical sensors, light-emitting diode, photovoltaics and cancer treatment.[50-56] In $\pi$-conjugated molecules, the overlap of the *p*-orbitals also imposes strong constraints that suppress bond rotation and thus lead to rigid backbones.[57-61] Polyacetylene (PA) is the simplest π-conjugated polymer (structure shown in Figure 3). In the COMPASS potentials, the dihedral potential is presented by a sum of three cosine functions (Equation 1). By comparing the potential parameters for PE and PA, much larger dihedral energy constants are found in PA (Table 2). The volume jump of PA at ~480 K shows that the phase transition temperature is ~80 K higher than that of PE (Figure 5a). Since PA and PE are of similar structure and similar inter-chain energy density, higher phase transition temperature in PA thus mainly comes from the stronger dihedral angle. PA dihedral angle distribution at 300K has a sharp mono peak, indicating all-*trans* conformations and straight chains (Figure 5b & upper inset). Although the overlap of dihedral angle distribution of PA decays as temperature increases, there is no sharp drop from 300K to 500K, indicating that no significant segmental rotation is present even above the ~480 K phase transition temperature. As a result, no sharp thermal conductivity drop is found during the ~480 K phase transition (Figure 3). Similar to Nylon, the volume expansion in PA around 480K is due to the collective change of the orientations of some chains in the nanofiber.

Table 2. Dihedral angle energy constants and pair energy densities of PE, PA, and PBT in COMPASS potential.

| Type of Dihedral Group | Leading Energy Constant (Kcal/mole) | Pair Energy Density (Kcal /Å$^3$/mol) |
|---|---|---|
| PE: C-C-C-C | 0.1223 | 0.01 |
| PA: C=C-C=C | 8.3667 | 0.01 |
| PBT: C=C-N=C | 6.8517 | 0.24 |
| N=C-C=C | 21.1715 | |
| C=C-S-C | 31.5576 | |
| N-C-S-C | 21.1715 | |

From Table 2, we see even larger dihedral energy constants in π-conjugated polymers with ring structures (*e.g.*, PBT) compared to those in PA. We simulated four π-conjugated polymer nanofibers with ring structures, including polythiophene (PT), poly(*p*-phenylene benzobisthiazole) (PBT), poly(*p*-phenylene) (PPP) and poly{diimidazo pyridinylene (dihydroxy) phenylene} (PIPD) (structures shown in Figure 3). The very large dihedral energy constants lead to sharp dihedral angle distributions at 300K for all four polymer nanofibers (Figure 5b). The dihedral angles at $0^o$ and $360^o$ indicate *cis* conformations, which come from the ring structures (Figure 5b & lower inset). Due to the existence of ring structures and the inter-chain π-π stacking,[62] large inter-chain energy density (Table 3) and steric effects stabilize the crystalline structure of these four π-conjugated polymer nanofibers. Therefore, no volume jump is observed from 300K to 500K for these polymer nanofibers (Figure 5a), and the dihedral angle distribution overlaps decay slower than PA (Figure 5c). Thermal conductivity calculations show that all five π-conjugated polymers have thermal conductivities higher than 20 W/mK at 300K (Figure 3). Unlike PE, the thermal conductivities of all the π-conjugated polymers only decrease gradually as temperature goes up, which leads to good thermal conductivities (15 ~ 50 W/Km) even up to 500K (Figure 3).

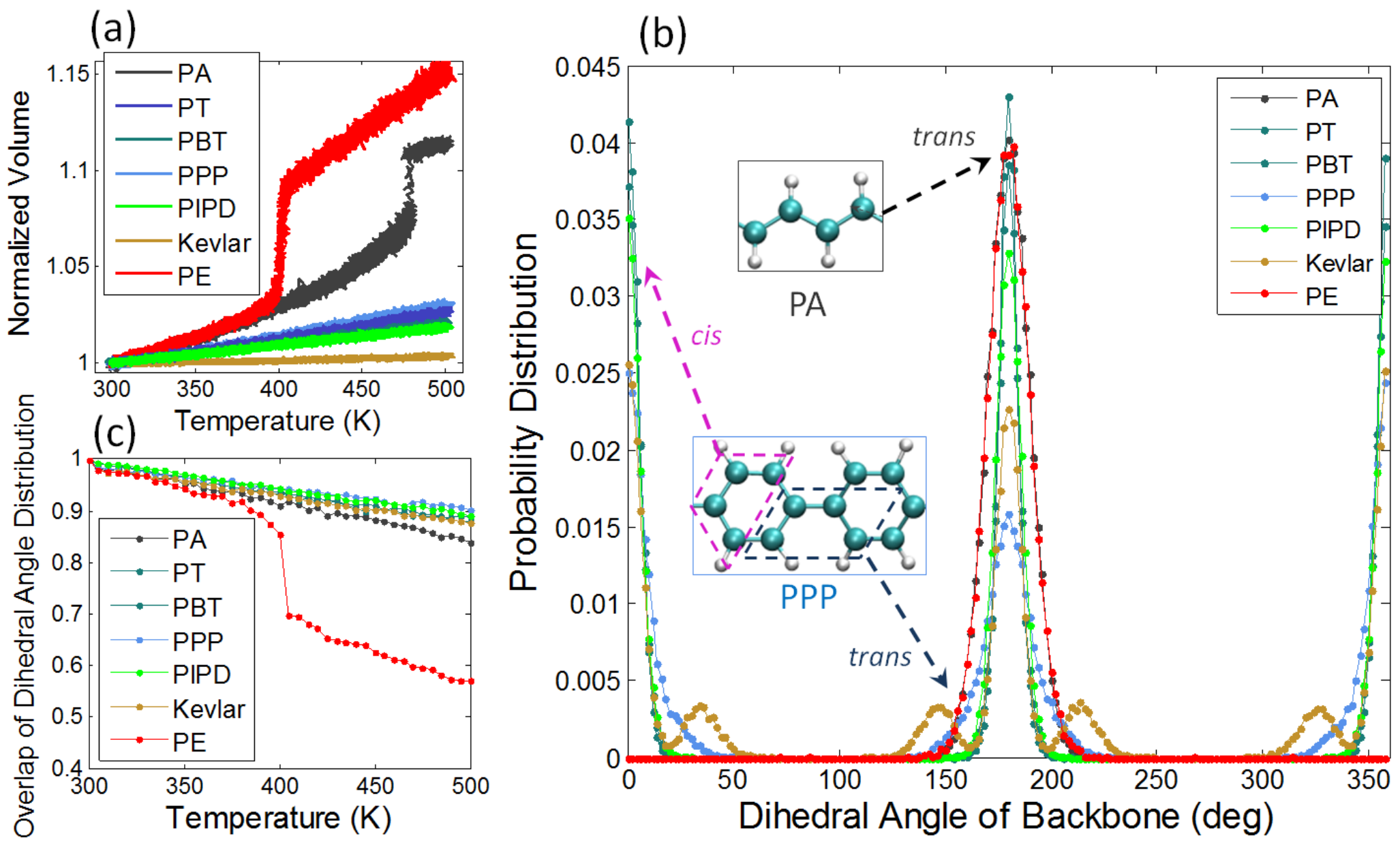

*Figure 5. Structural characteristics of polymer nanofibers with different backbone stiffness: (a) volume from 300K to 500K, (b) dihedral angle distribution at 300K, and (c) the overlap of dihedral angle distribution from 300K to 500K for PE, PA, PT, PBT, PPP, PIPD, and Kevlar.*

Table 3. Pair energy density, phonon group velociy, volumetric heat capacity, phonon mean free path, and thermal conductivity for the ten folymer nanofibers studied.

| | Pair Energy Density (Kcal/Å$^3$/mol) | Phonon Group Velocity (m/s) (averaged) | Volumetric Heat Capacity (J cm$^{-3}$ K$^{-1}$) | Mean free path (nm) | Thermal Conductivity at 300K (W/mK) |
|---|---|---|---|---|---|
| PPP | 0.19 | 7663.86 | 4.30 | 1.37 | 45.13 |
| PA | 0.01 | 7410.13 | 3.37 | 4.00 | 100.09 |
| PBT | 0.24 | 7241.58 | 3.90 | 1.34 | 37.72 |
| PIPD | 0.16 | 7074.79 | 4.26 | 0.75 | 22.42 |
| PT | 0.06 | 7069.99 | 2.93 | 3.04 | 62.71 |
| PE | 0.01 | **5292.94** | 4.89 | 1.90 | 49.2 |
| PK | 0.38 | 4264.8 | 4.40 | 0.86 | 16.08 |
| Nylon | 0.07 | 3566.04 | 4.89 | 0.46 | 7.99 |
| Kevlar | 0.25 | 3230.32 | 4.16 | 0.60 | 8.05 |
| Teflon | 0.74 | 2801.78 | 4.43 | 0.28 | 3.43 |

We found that the $\pi$-conjugated nature is important to achieving high thermal stability and high thermal conductivity at the same time. One important feature of $\pi$-conjugated polymer is the existence of alternating single and double bonds. Kevlar also has benzene rings that have alternating single and double bonds, but the delocalization of *p*-electrons over the whole molecule is not possible due to the presence of the carboxamide group (portion highlighted in blue, Figure 3). Such a characteristic makes Kevlar not a $\pi$-conjugated polymer. Figure 5 shows that there is no phase transition from 300K to 500K and no significant random segmental rotations due to the large pair energy density in Kevlar (Table 3). However, the thermal conductivity of Kevlar is below 10 W/mK in the whole temperature range, which is at least 50% smaller than the conjugated polymers. This finding is consistent with a recent measurement which shows that Kevlar has much lower thermal conductivity compared to $\pi$-conjugated polymer nanofibers.[6] We found that segmental rotation is intrinsic in Kevlar chains at 300K (Figure 5b, smaller yellow shoulder peaks). However, unlike Teflon which has intrinsic segmental rotations that are very randomized, long range order is still preserved in the rotated Kevlar segments. Therefore, the low thermal conductivity of Kevlar cannot be explained by the rotational disorder-phonon scattering mechanism. We attribute this to the characteristics of the phonons which are analyzed in the following section.

## 2.4. Phonon Dispersion Relation Analysis

The very high thermal conductivity of PE has been well studied by both simulations and experiments.[2, 3, 8] The high thermal conductivities in $\pi$-conjugated polymers, including PA and PPP, have also been reported by studying stand-alone single chains by simulation.[63] Experimental data also show that the thermal conductivities of $\pi$-conjugated polymer nanofibers (PIPD and PBT) are much higher than those of other polymers (*e.g.* Kevlar and Vectra).[6] Here, to explain the higher thermal conductivities in $\pi$-conjugated polymer nanofibers, the following equation for thermal conductivity of one-dimensional systems is analyzed:

$$\kappa = c \cdot \bar{v} \cdot \bar{l} \tag{3}$$

where $\kappa$ is thermal conductivity along polymer chains, $c$ is volumetric heat capacity, $\bar{v}$ is the average phonon group velocity, and $\bar{l}$ is the phonon mean free path. Since our MD simulations are in the classical limit, the volumetric heat capacity can be expressed as $c = 3k_B\frac{N}{V}$, where $k_B$ is Boltzmann constant, $N$ is the number of atoms, and $V$ is volume (Table 3). Although the volumetric heat capacity varies from 2.93 J/(cm$^3$K) to 4.89 J/(cm$^3$K) for the ten polymers, the smallest values are found in PT (2.93 J/(cm$^3$K)) and PA (3.37 J/(cm$^3$K)) – the two $\pi$-conjugated polymers with the highest thermal conductivities in all ten nanofibers. Therefore, volumetric heat capacity is not the key to the high thermal conductivity in these nanofibers.

Another factor is the phonon group velocity. To calculate the phonon group velocity,

phonon dispersion relation is first calculated by performing two-dimensional Fourier transforms of the velocities of the backbone atoms obtained from MD simulations (Figure 6, see Appendix B for calculation details).[64-66] Since we believe that acoustic phonons are the dominant heat carriers that contribute to thermal transport, we have only included the low frequency regions including acoustic phonons in Figure 6 for clearer comparison. The complete phonon dispersion relations are shown in Appendix B. For each acoustic branch, the phonon group velocity is estimated as the slope between the original and the point of the max frequency (blue line in PPP, Figure 6). The average phonon group velocity is approximated as the arithmetic mean of all four acoustic branches[2] (one longitudinal, two transverse, and one torsional branch, highlighted by red dots in Figure 6), and the results are listed in Table 3. Although such estimation is relatively rough, it gives us a general idea of the relative magnitude of the phonon group velocities of different polymer fibers so that a comparison among them can be made. It is clear that all the π-conjugated polymers have mean phonon group velocities above 7000 m/s (Table 3, red) – higher than those of the other nanofibers.

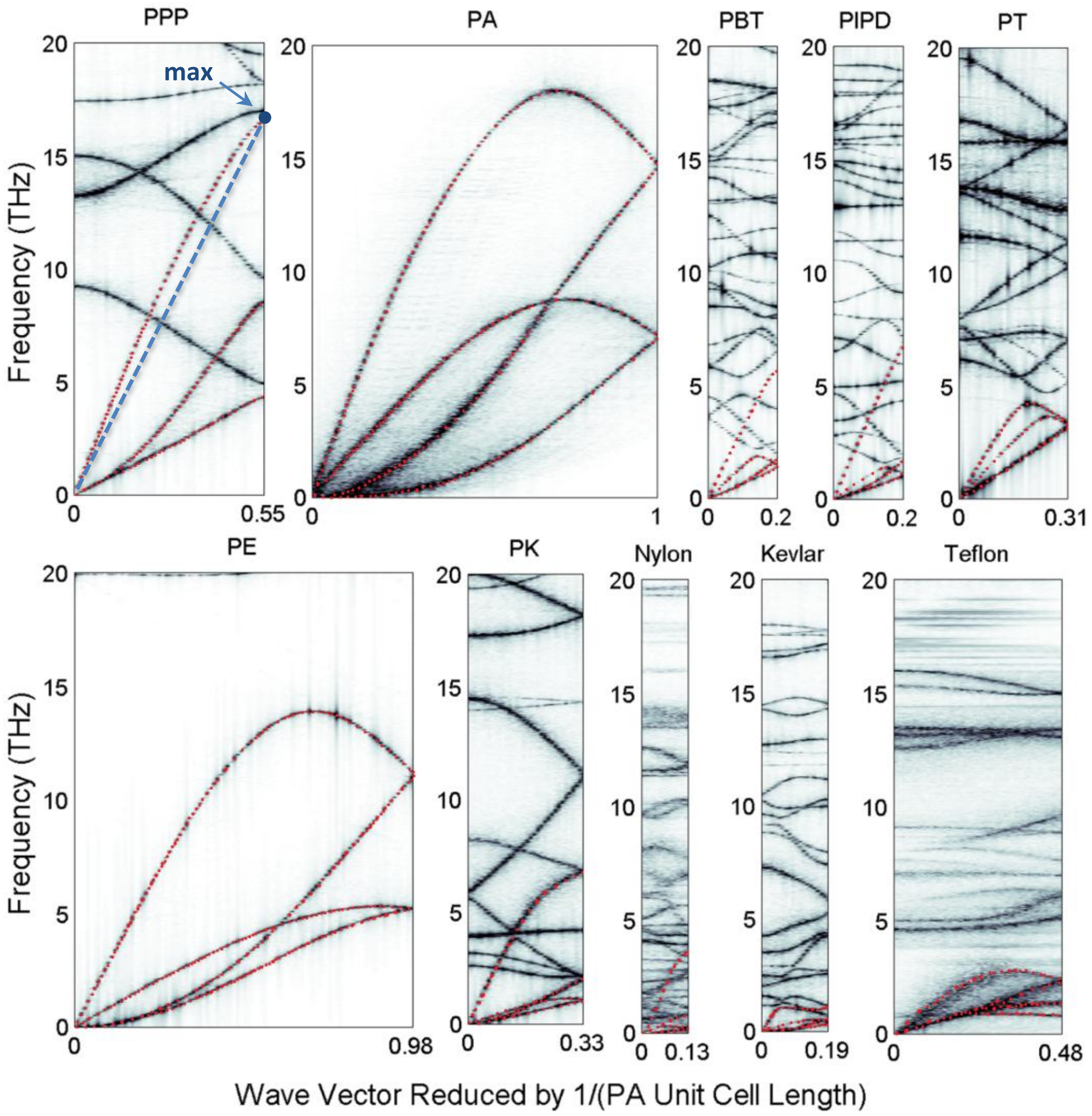

*Figure 6. Phonon dispersion relation for the ten polymer nanofibers.The acoutic phonon branches are marked with red dots. The average phonon group velocity of one branch is estimated as the slope between the origin and the point of the max frequency (e.g., blue line, in PPP).*

At the molecular level, strong bonds and small atomic masses can usually enable high phonon group velocities. In the COMPASS potential, bond strength is defined by the energy constants in the following formula:

$$E_b = K_2(r - r_0)^2 + K_3(r - r_0)^3 + K_4(r - r_0)^4 \quad (4)$$

where $r$ is the distance between the two bonded atoms, $r_0$ is equilibrium bond distance, $K_2$, $K_3$, and $K_4$ are the energy constants defining the strength of the bond. Table 4 shows that larger bond strength is found in π-conjugated polymers due to *p*-orbitals conjugation and delocalized electrons. The carboxamide group (in Nylon and Kevlar), on the other hand, will decrease the strength of the bond between carbon and carbonyl carbon in the backbone, leading to lower phonon group velocities. As a result, Nylon and Kevlar have low phonon group velocities (Table 3). Compared to light hydrogen atom, oxygen and fluorine atoms are much heavier, which can also explain the relatively low phonon group velocity in PK, Teflon, and Kevlar.

Table 4. Typical bond parameters in PE, PA, PBT, and Nylon

| | $r_0$ | $K_2$ | $K_3$ | $K_4$ |
|---|---|---|---|---|
| PE ( $H_2C$—$CH_2$ ) | 1.5330 | **299.6700** | **-501.7700** | **679.8100** |
| PA ( HC⩶CH ) | 1.4170 | 470.8361 | -627.6179 | 1327.6345 |
| PBT ( C⩶NH ) | 1.3485 | 508.8587 | -977.6914 | 1772.5134 |
| Nylon( $H_2C$—CONH ) | 1.5202 | 253.7067 | -423.0370 | 396.9000 |

With the thermal conductivity, volumetric heat capacity and phonon group velocity, we can calculate the effective phonon mean free path at 300K according to Equation 3. Due to the approximations used in calculating volumetric heat capacity and phonon group velocity, the mean free paths calculated from Equation 3 are not expected to be exactly accurate. However, the comparison of the phonon mean free path among the fibers can offer valuable insights to the phonon transport physics in these nanofibers. For example, Table 3 clearly shows that polymer nanofibers with good along-chain orders, such as the π-conjugated polymers and PE, tend to have larger phonon mean free paths. As discussed previously, the segments in the π-conjugated polymers are well ordered and thus the disorder-phonon scattering is weak, enabling long phonon mean free paths. It is also interesting to find out that the effective phonon mean free path of Teflon is the smallest. This is believed to be related to the intrinsic segmental disorders in the Teflon chains which scatter phonons.

## 3. Conclusion

Due to the overlap of the *p*-orbitals in π-conjugated polymers, rigid backbones in PA, PIPD, PBT, PPP and PT suppress segmental rotations and thus phonon scattering. The π-conjugation also enables strong backbones and thus large phonon group velocities. Therfore, π-conjugated polymers are found to have both good thermal stabilities and high thermal conductivities. Strong inter-chain interaction in PK, Nylon, Teflon and Kevlar can suppress thermal expansion and thus stabilize the structures at high temperatures. However, the presence of heavier atoms (*e.g.* oxygeon) and weaker backbone bonds will lead to low phonon group velocities and thus low thermal conductivities of these polymers. Polymers with intrinsic segmental rotations (*e.g.*, Teflon) will not have high thermal conductivities due to the disorder-phonon scattering. Such a factor also lead to the significant reductions in thermal conductivities of nanofibers which attain segmental rotations at high temperatures. This work made detailed connections between the molecular characteristics and macroscopic thermal properties of polymer nanofibers. *It provides a general rule of designing thermally conductive polymer nanofibers with high temperature stabilities: one should choose molecules with intrinsically ordered backbones, strong backbone bonds and strong dihedral angles.* According to this rule, π-conjugated polymers are singled out as a category of polymers that are ideal for fabricating nanofibers with high thermal conductivities for high temperature applications.

## 4. Simulation Section

In the present work, MD simulations with the COMPASS are used to investigate the thermal stabilities and thermal conductivities of 10 different polymer nanofibers, including PA, PT, PPP, Kevlar, Nylon 6-6, PE, PBT, PK, PIPD, and Teflon. The COMPASS potential was created to accurately simulate the structural, vibrational, and thermo-physical properties of polymeric materials in isolated and condensed phases.[37, 38] It is thus chosen for the present study in which the morphology-thermal property relation is important. The initial fiber structures are constructed based on previous literatures (see Appendix A), and then minimized using multiple optimization steps, including steepest descent, conjugate gradient, and Newton minimizations, to guarantee reaching the lowest energy morphology. Periodic boundary conditions in all three spatial dimensions are applied to avoid surface effects.[67] This corresponds to the thick fiber limit where surface effect is not important. A fourth periodic condition which links one end of a chain to the other end of its image is used to simulate infinitely long chains, which mimic the monodisperse ultrahigh molecular weight polymers.[68] The number of chains are chosen so that the cross sectional dimensions are larger than two times of the interaction cutoff (10 Å), avoiding the double counting of the interactions between atoms.

Previous studies have shown that thermal conductivities of polymers chains in the chain length direction can be a function of domain length due to boundary phonon scattering.[9, 11, 63] In order to make fair comparison among different nanofibers, the nanofibers are constructed by controlling the number of segments in polymer chains so that the simulation domain lengths are all around 50 nm at 300K (see Appendix A). From the minimized structures, the systems are simulated in NPT ensembles for 1 ns to obtain the stable nanofiber structures at different temperatures and 1 atm before any further calculations or characterizations are performed. Such a step effectively eliminates large stresses in the system, especially in the along-chain direction.

Non-equilibrium molecular dynamics (NEMD) simulation is used to calculate the thermal conductivity of different nanofibers.[3, 9-11, 69] In this method, a temperature gradient is created and maintained by Langevin thermostats controlling the temperatures of the two ends of the simulation domain. The temperatures of the heat sink and heat source regions are set to 15 K lower and higher than the average system temperature, respectively. After steady state is reached, temperature gradient $(dT/dx)$ is obtained by fitting the linear portion of the temperature profile, and heat flux can be calculated using $J = dQ/dt/S,$ where $dQ/dt$ is the average energy change rates of the two Langevin thermostats, and $S$ is the cross sectional area. The thermal conductivity is then calculated by Fourier's law, $\kappa = -J/(dT/dx)$. Figure 7 shows a representative setup and temperature profile of NEMD. For each simulation, four thermal conductivity values are calculated for different time blocks in the steady state, and the final value is the average of them with the error bar being the standard deviation.

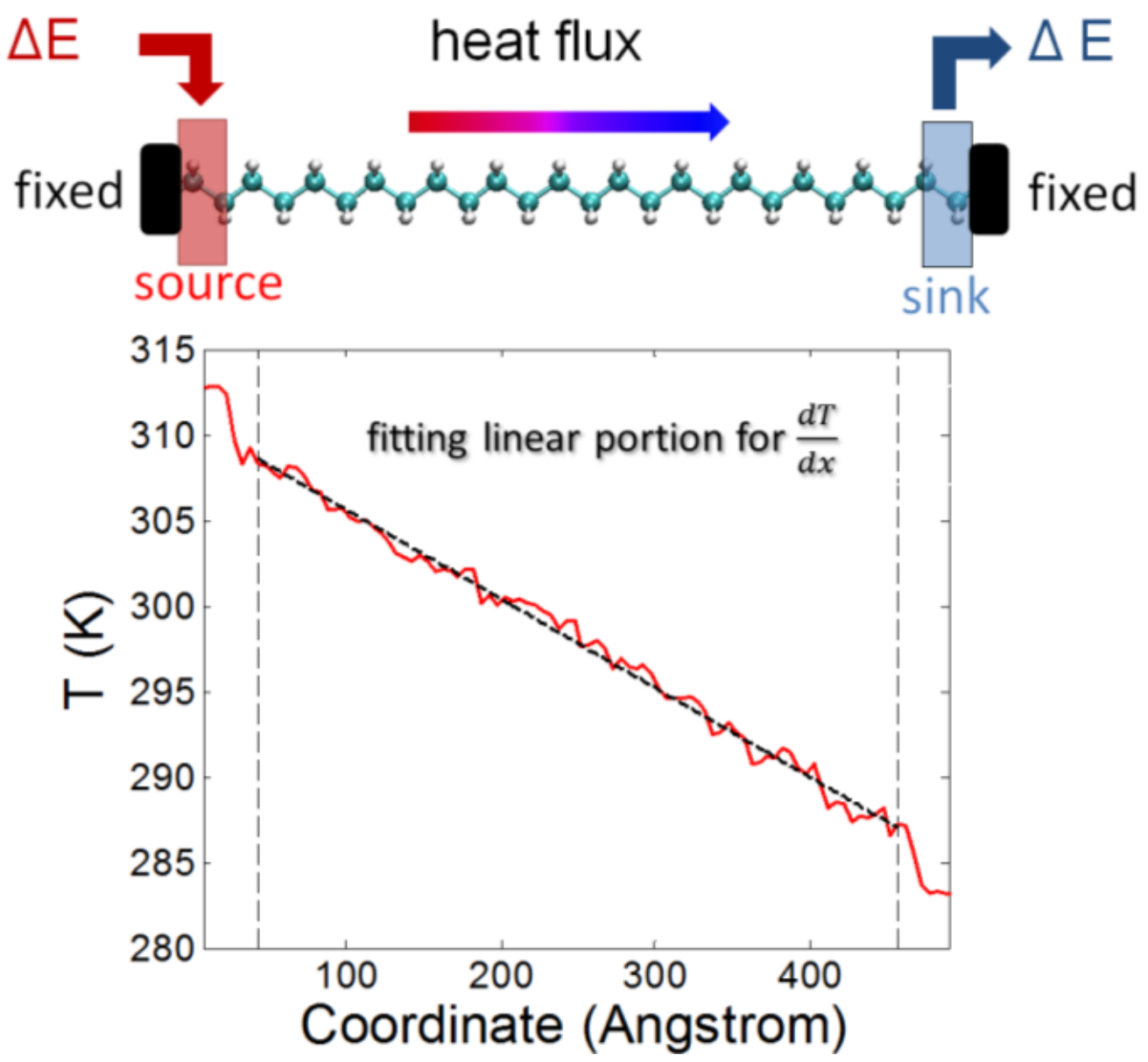

*Figure 7 NEMD scheme and corresponding temperature profile.*

All the simulations are carried out using the large-scale atomic/molecular massively parallel simulator (LAMMPS).[70] The time step is chosen to be 0.25 fs, and a 10 Å cutoff is used for the 6-9 Lennard-Jones interaction, allowing for chain-chain interactions between the third and the fourth nearest-neighboring chains.

## Appendix A: Polymer Nanofiber Structures Construction

The initial structures of ten different polymer nanofibers are built based on the structures documented in literatures.[3, 71-79] The polymers studied include PA,[72] PT,[73] Kevlar,[75] PIPD,[71] PPP,[74] Nylon 6-6 (imported from Material Studio), PE,[3] Teflon,[76] PBT,[77] and PK (imported from Material Studio) (Table A1). All the initial structures are minimized using multiple algorithms, including Steepest Descent, Conjugate Gradient, and Newton minimization (Smart Minimizer, Discover module of Material Studio, Accelrys Inc.), before they are used for MD simulations. Runs in NPT ensembles for 1 ns are performed to obtain the stable structures before the thermal conductivity calculations at different temperatures from 300K to 500K. The structures of crystalline polymer nanofibers can be a strong function of temperature, drawn ratio, and pressure,[16, 80-84] and thus only the supercell dimensions of the stable structures at 300K and 1 atm are listed (Table A2).

Table A1. Structures and dimensions of the unit-cell constructed for different polymer nanofibers.

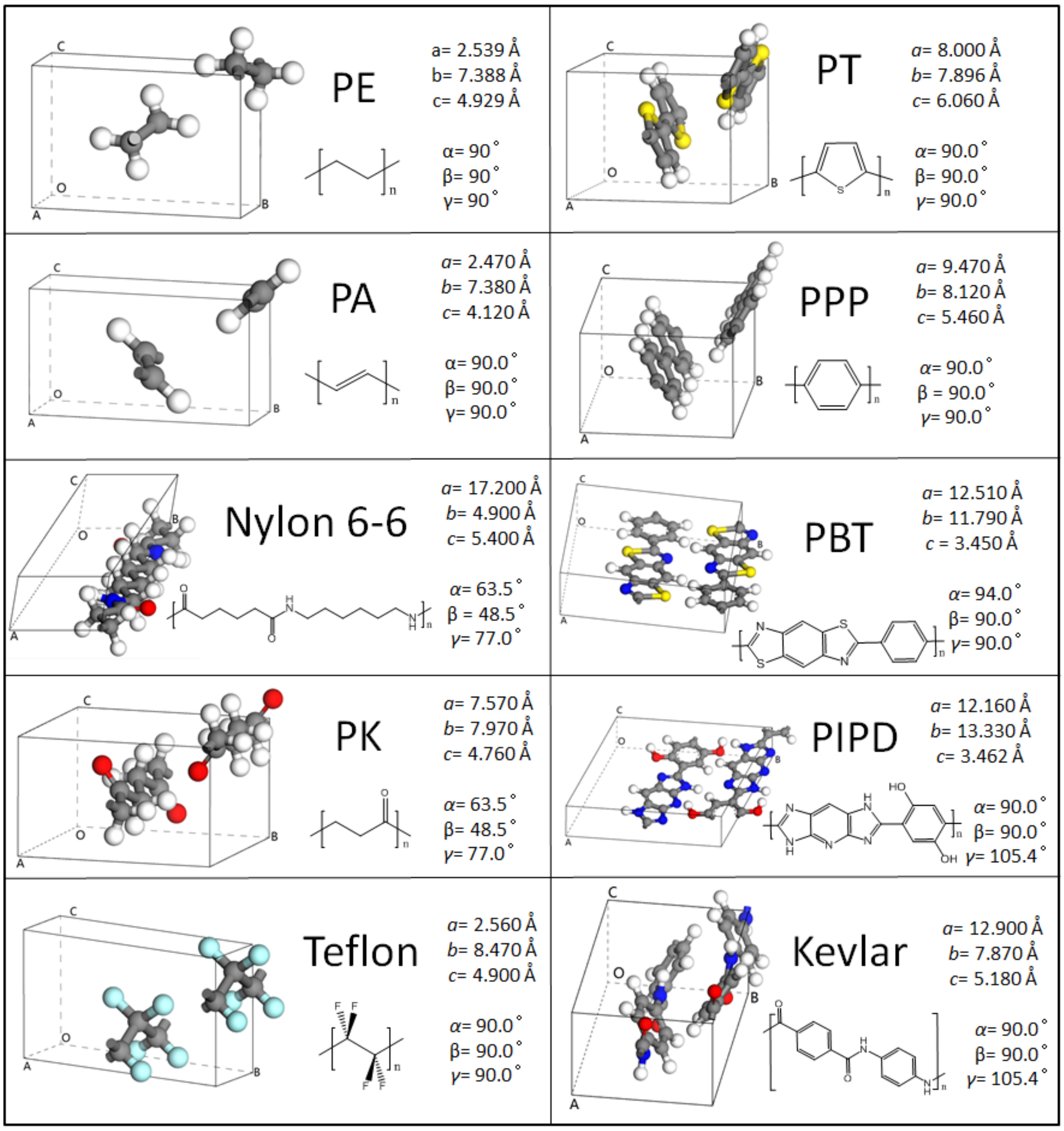

● α = ∠ BOC, β = ∠ AOC, γ= ∠ AOB

Table A2. The construction and dimensions of the super-cell for different polymer nanofibers.

| | Segments per Chain | Chain Number | Super-cell Dimensions (300K and 1 atm) | | |
|---|---|---|---|---|---|
| | | | X (Å) | Y (Å) | Z (Å) |
| PE | 400 | 40 | 506.70 | 24.12 | 33.15 |
| PA | 400 | 56 | 508.07 | 34.82 | 27.19 |
| Nylon 6-6 | 29 | 81 | 501.76 | 43.33 | 34.75 |
| PK | 134 | 40 | 496.65 | 27.98 | 29.03 |
| Teflon | 400 | 56 | 445.64 | 37.45 | 37.74 |
| PT | 160 | 40 | 562.39 | 32.28 | 29.87 |
| PPP | 110 | 40 | 471.91 | 34.14 | 26.28 |
| PBT | 40 | 54 | 478.28 | 35.75 | 32.17 |
| PIPD | 42 | 54 | 496.81 | 36.37 | 32.93 |
| Kevlar | 40 | 40 | 506.70 | 24.12 | 33.15 |

- super-cell dimension: X ×Y ×Z, with X parallel to OA, YZ in cross sectional plane

## Appendix B: Phonon Dispersion Relation Calculation

With the initial unit cell structures, single polymer chains are constructed with the same number of segments as in the nanofibers (Table A1 & A2). For each simulation, only one isolated polymer chain is placed in the supper cell, and the cross sectional dimensions are set to be enough large (50 Å×50 Å), preventing the interaction with its own image. After the minimization, these single chains are first simulated at 2 K with fixed volume, and then simulated in ensembles of constant volume and energy (NVE) for 50 ps. During one NVE run, the velocity of every backbone atom is recorded every 5 fs in a two-dimensional matrix. With the velocity matrixes from MD simulations, two-dimensional Fourier transform of the atomic velocity of one certain atom in the unit cell is performed:

$$\Phi(\omega, k) = \sqrt{\sum_{\alpha}^{3} \left| \frac{1}{N} \sum_{n=0}^{N-1} e^{i\frac{n}{N}k} \int v_{\alpha}(n,t) e^{-i\omega t} dt \right|^{2}}, \ \alpha = x, y, z \qquad \text{(B1)}$$

where $v_{\alpha}(n,t)$ is the atomic velocity, $\omega$ the frequency, $k$ is the wavevector, $n$ is the index of repeating unit along chain direction, and $N$ is the number of the repeating unit. Then the phonon dispersion relation is calculated by averaging the two-dimensional Fourier transform results of all the backbone atoms in the unit cell (Figure B1).

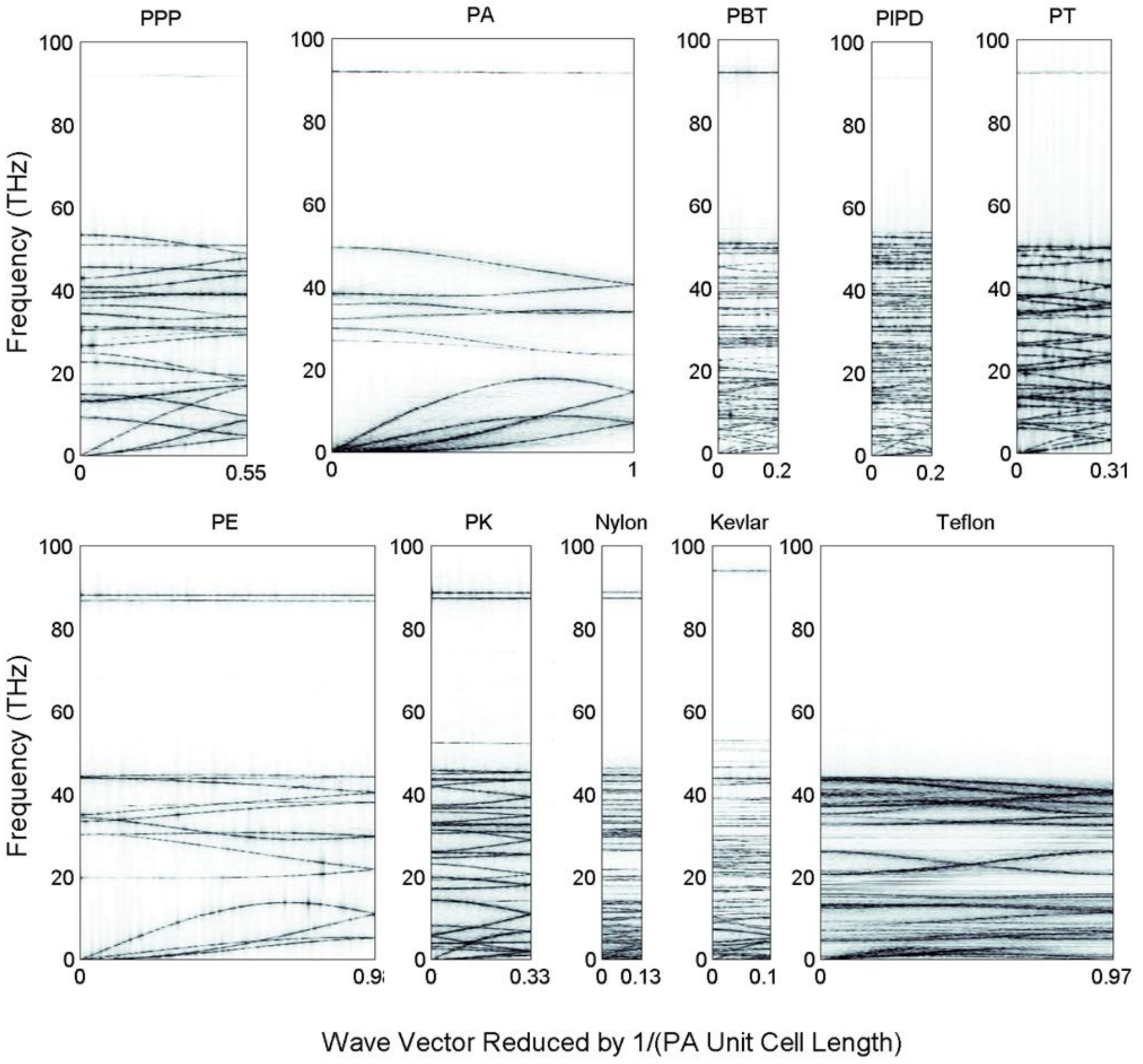

Figure B1 *Phonon dispersion relation for the ten polymer nanofibers.*

## Acknowledgements

*This research was supported in part by the Notre Dame Center for Research Computing and NSF through TeraGrid resources provided by SDSC Trestles under grant number TG-CTS100078. The authors acknowledge the financial support from American Chemistry Society (PRF# 54129-DNI10) and the startup fund from the University of Notre Dame.*

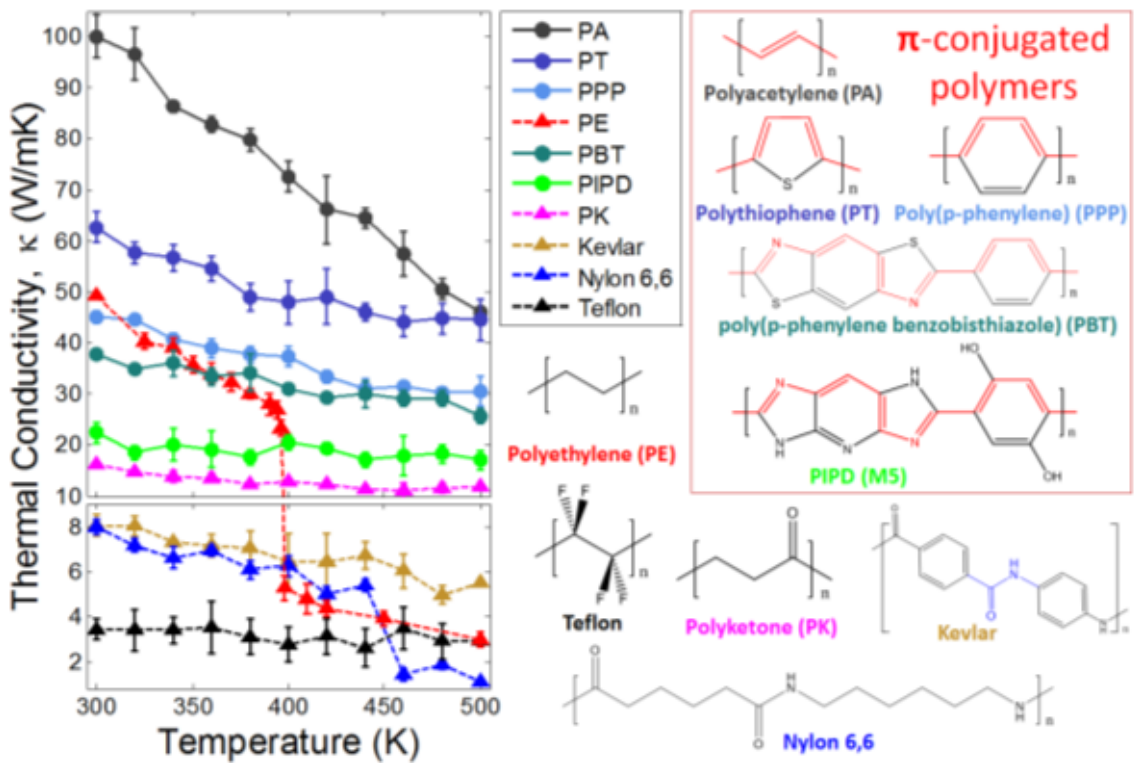

Table of Contents